# Search for determining the role of various physical properties of the fusion reactions in the proximity formalism

**Gharaei, Reza; Sarvari, Elham**

*Department of Physics, Hakim Sabzevari University, Sabzevar*

**Abstract:** *This work is focused on the analysis of the corrective effects of the temperature, surface tension, and nuclear matter density on the fusion barriers and also fusion cross sections caused by the original version of the proximity formalism (Prox.77 model) for 62 fusion reactions with conditions of $65 \leq Z_1 Z_2 \leq 1520$. A systematic comparison between the theoretical and empirical data of the height and position of the barrier as well as the fusion cross sections reveals that the agreement of these data improves by imposing each of the mentioned physical effects on the Prox.77 model for our selected mass range. Moreover, it is shown that the density-dependence effects have the most important role in improving the theoretical results of the proximity potential.*

## 1. Introduction

One of the complexities in the era of analysis of the fusion process in dual-core targets and projectiles is understanding the strength of the interactions between these two cores from the moment of their overlap in the input channel until the formation of the compound nucleus in the output channel. It has been long since comprehensive knowledge of nuclear forces has been one of the most challenging topics in the study of fusion reaction.

However, our understanding of the Coulomb confrontations between two interacting nuclei and the estimate of their strength during the fusion process is complete. In recent decades, various theoretical models for calculating the nuclear potential in fusion reactions have been introduced, including the proximity force theorem [1], the double-folding model [2], and Skyrme's density and energy-dependent forces [3].

For the moment, the proximity potential formalism is one of the models that are efficient and also does not have much analytical complexity. The results of studies reveal that despite the acceptable accuracy of the formalism in reproducing experimental data include various fusion reactions, the results of the proximity potential model still need to be improved [5,4].

Studies such as [7-9] have analyzed the proximity potential by applying effects such as the temperature of the compound nucleus, surface tension, and nuclear material density.

So far no systematic studies have been operated to simultaneously examine the role of each of these effects in proximity formalism and also to determine the most effective feature known to enhance the results of this perspective. So the present study is dedicated to the analysis of the potential barrier as well as the fusion cross-section resulting from the original version of the proximity formalism, Proximity 1977 (Prox.77), before and after applying these effects for 62 fusion reactions.

## 2. Formulation

All proximity potentials are based on the proximity force theorem. According to which, "the force between two gently curved surfaces in close proximity is proportional to the interaction potential per unit area between the two flat surfaces". So according to the proximity force theorem, we want to go into as much detail as possible about the associated formulas in the Prox.77 model to calculate the fusion barrier, and knowing about how to apply the correction effects associated with this model.

### 3. Proximity potential theorem and Prox 77

According to the original version of proximity, when the surface of two interacting nuclei reaches a distance of 2-3 fm from each other, a force will appear between them, which is called the "proximity force" [1]. In 1977, a group of scientists used this theory to propose a model for calculating the potential of a nucleus. So based on this theory, the interaction potential $V_N(r)$ between two surfaces can be written as

$$V_N(r) = 4\pi \bar{R} b \gamma \Phi(\xi = r - c_1 - c_2) \quad \text{MeV} \quad (1)$$

Where, $\bar{R}$ and b are the reduced radius of the target and projectile system and the surface thickness, respectively [4,1]. In addition, γ is the surface energy coefficient taken from the Lysekil mass formula (in $MeV/fm^2$), which depends on the symmetry or asymmetry of the nuclei in terms of the number of protons and neutrons.

$$\gamma = \gamma_0 \left[1 - k_s \left(\frac{N-Z}{N+Z}\right)^2\right] \quad (2)$$

In this formula, $A_s = (N-Z)/(N+Z)$ is called the asymmetry parameter and also additionally, in the Prox.77 model, the coefficients $\gamma_0$ and $k_s$ are called surface energy constant and surface-asymmetry constant, which have the values of 1.01734 and 1.79 MeV/$fm^2$, respectively.

### 4. Corrective effects of temperature

In this paper, to apply temperature corrections to the approximate formalism, we used the generalized form of surface energy coefficient γ as follows [6],

$$\gamma = \gamma_0 \left(1 - k_s \left(\frac{N-Z}{A}\right)^2\right) \left(2 - \frac{T}{T_B}\right)^{1.5} \quad (3)$$

Where, T and $T_B$, are the temperatures corresponding to the energy of the center of mass $E_{c.m.}$ and the energy of the Coulomb barrier $E_B$, respectively. The calculation details are mentioned in reference [7].
In this study, we named the temperature-dependent proximity potential Prox.77 (TD).

### 5. Corrective effects of surface tension

According to the study in 2010, the role of the effects of surface energy coefficient γ on fusion barriers from the Prox.77 potential model was evaluated by selecting different values of $\gamma_0$ and $k_s$ constants [8]. Based on the results if somebody used the set of values, $\gamma_0 = 1.460734$ and $k_s = 4$ to calculate the coefficient γ in equation (2) will be able to get the best results to compare to experimental data for the fusion barrier heights and position in computation with the other selected value sets.

We name the results of the modified proximity potential through the effects of dependence on the surface energy coefficient γ with Prox.77 (GD).

## 6. Corrective effects of density-dependence

In general, to apply such effects to the fusion process, we can use the double-folding (DF) model associated with density-dependence interactions M3Y. In a recent study, these effects have been indirectly considered in the proximity model by analyzing the behavior of the universal function at radial distances around the Coulomb barrier [9]. Accordingly, a new form for the universal function introduced, which is as follows:

$$\Phi(\xi) = \frac{P_1}{1 + e^{\frac{\xi + P_2}{P_3}}} \quad (4)$$

So, the values obtained for the constants $p_1$, $p_2$, and $p_3$ are -17.72, 1.30, and 0.854, respectively. Note that we named the result for a modified version of proximity potential, Prox.77 (DD).

## 7. Discussion and Results

The present study investigates the role of three different physical effects on the characteristics of the fusion barrier (its height and location) as well as the fusion cross-section from the proximity-based potential. To achieve this goal, we have selected 62 different fusion reactions, which are selected $60 \leq Z_1Z_2 \leq 1520$ for the multiply of the atomic numbers of the target nuclei and their projectile.

On the other hand, we assumed that the selected nuclei are all spherical in their ground state. In the first step, we are interested in evaluating the behavior of the total interaction potential at different radial distances. Based on an arbitrary theoretical model with a simple form of Coulomb potential as $V_C(r) = Z_1Z_2e^2/r$, we are able to calculate the total interaction potential $V_T(r)$ at the distance between the surfaces of the interacting pair of nuclei.

Fig. 1 shows the behavior of the total interaction potential in terms of radial distance r based on the original version of the proximity potential as well as its modified versions for an arbitrary fusion reaction such as $^{16}O+^{40}Ca$.

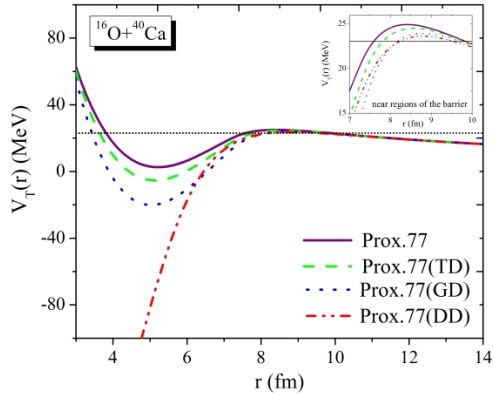

*FIG. 1. Total interaction potential behavior in terms of radial distance (r) between the surfaces of the interacting pair of nuclei based on Prox.77, Prox.77 (TD), Prox.77 (GD), and Prox.77 (DD) potential models for fusion reaction $^{16}O+^{40}Ca$.*

As can be seen from Fig. 1, each of the effects includes temperature, surface energy coefficient and nuclear matter density-dependence on the Prox.77 model reduces the height as well as increases the potential depth in the interior region.

In fact, after modifying the proximity formalism through each of the mentioned effects, the resulting barrier height is 0.42, 1.00, and 1.32 MeV, respectively, and they get closer to the experimental value $V_B^{Emp} = 23.06$ MeV.

It should be noted that the absence of valleys in the interior region of the potential in the modified version of Prox.77 (DD) can be attributed to the characteristics of the DF model, which this modified model originates.

To better understand the role of the correction effects study on the proximity formalism in the entire selected mass range in this work, we have calculated the percentage difference between the theoretical and experimental values of height and position of the Coulomb barrier for a total of 62 fusion reactions. Which is as follows:

$$\Delta X(\%) = \frac{X^{Theor.} - X^{Emp.}}{X^{Emp.}} \times 100 \tag{5}$$

Where $X = R_B$ or $V_B$. The results of calculations for the values of $\Delta R_B$ (%) and $\Delta V_B$ (%) as a function of $Z_1 Z_2$ are shown in parts (a) and (b) of Fig. 2, respectively.

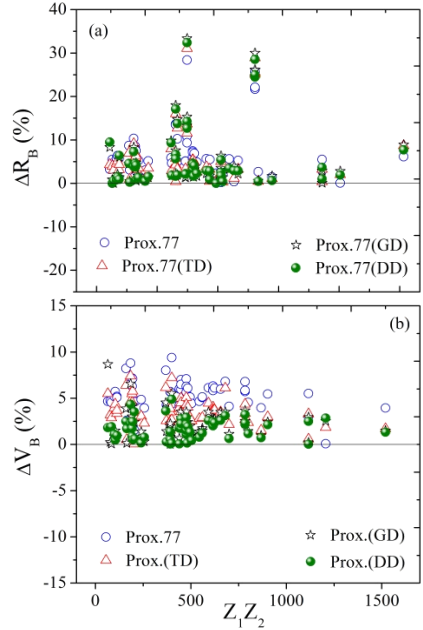

FIG. 2. The values of a) $\Delta R_B$ (%) and b) $\Delta V_B$ (%) in terms of multiplication of $Z_1 Z_2$ based on the original version of the proximity potential and its three modified versions within the selected mass range mentioned above.

The results of fig. 2 show that the mean deviations for the barrier height from the Prox.77, Prox.77 (TD), Prox.77 (GD) and Prox.77 (DD) models are equal to 5.621%, 3.506%, 2.198% and 1.657%, respectively.

For fusion barrier, the mean deviations are 6.516%, 5.869%, 5.575% and 5.389%, respectively. With a closer look at these values, it reveals that applying each of the physical effects such as temperature, surface energy coefficient, and density-dependence to the Prox.77 model can improve agreement between the theoretical and experimental values of fusion barrier height and position in our selected mass range.

In addition, it is observable that among the above-mentioned effects, the density dependence has the greatest contribution in reducing the difference between theoretical and experimental values of $R_B$ and $V_B$.

Another factor that is generally research evaluates the process of variation in the theatrical study of fusion reactions is fusion cross-section $\sigma_{fus}$. In the present work, a one-dimensional barrier-penetration model is used to calculate the values of this quantity, the analytical details are described in reference [10].

The results reveal that the effect of each correction factor to the Prox.77 model increases the theoretical values for the fusion cross-section and make them closer to the corresponding empirical values in heavy ions collision systems.

On the other hand, it is proved that density-dependent factors have the greatest effect on the agreement between theoretical and experimental values of $\sigma_{fus}$. The energy-dependent behavior of this quantity in terms of the center of mass-energy in Fig. 3 confirms the above results for the two arbitrary fusion reactions in $^{24}Mg+^{35}Cl$ and $^{16}O+^{116}Sn$.

## 8. Summery

In the present work, we systematically study the effectiveness of the proximity potential model with three important physical effects of the interactional dual-core fusion channel. Meanwhile, the temperature effect, surface tension, and density-dependence effects in the selected mass range of $65 \leq Z_1Z_2 \leq 1520$.

Another approach in this study is to introduce the most important physical factor that has the greatest effect on the results of the Prox.77 model. Also, the results show that applying each of these effects individually improves the $R_B$, $V_B$, and $\sigma_{fus}$ values based on the proximity formalism in our selected mass range. Moreover, we also have shown that the density-dependence effects have the most important role in improving the theoretical results of the proximity potential.

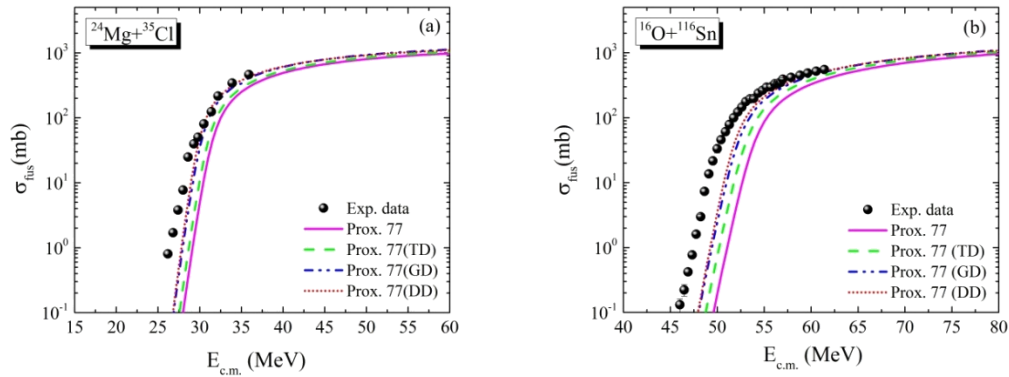

FIG 3. *Energy-dependence behavior of theoretical and experimental cross-section values based on four potential models Prox.77, Prox.77 (TD), Prox.77 (GD) and Prox.77 (DD) for two fusion reactions a) $^{24}Mg+^{35}Cl$ and b) $^{16}O+^{116}Sn$. The experimental data of these two reactions were extracted from references [11] and [12], respectively.*